\def\simt0{\mathrel{\mathop{\sim}\limits_{{}^{\tau\to 0}}}}
\def\tot0{\mathrel{\mathop{\longrightarrow}\limits_{{}^{\tau\to 0}}}}
\documentstyle[12pt]{article}

\begin{document}

\title{ Bosonization and even Grassmann variables}
\author{M.B. Barbaro$^a$, A. Molinari$^a$ and F. Palumbo$^b$
\thanks{This work is carried out in the framework of the
European Community Research Program "Gauge theories,
applied supersymmetry and quantum gravity" with a financial
contribution under contract SC1-CT92-0789 .}
\\
\\
\em $^a$ Dipartimento di Fisica Teorica dell'Universit\`a
 di Torino and \\
\em Istituto Nazionale di Fisica Nucleare, Sezione di Torino, \\
\em via P.Giuria 1, I-10125 Torino, Italy \\
\em $^b$ INFN - Laboratori Nazionali di Frascati, \\
\em P.O.Box 13, I-00044 Frascati, Italy}
\date{July 1996}
\maketitle
\begin{abstract}
We test a new approach to bosonization in relativistic field theories and
many-body systems, based on the use of fermionic composites as integration
variables in the Berezin integral defining the partition function of the
system. The method appears promising since at zeroth order it correctly
describes the propagators of the composites, which can 
be evaluated in a number of significant cases. 
Still to be established is a general procedure for deriving the free action of
the composites starting from the one of the constituents. 
To shed light on this problem and to explore further features of the method we
study a simplified version of the BCS model. In this simple case 
the action of the composites can indeed be obtained: whether this result
can be generalized it remains however to be seen.
Yet an interesting property of the wave operators
appearing in the free actions of bilinear composites already emerges from the
simple problem we have treated: here the wave operators do not describe any time
evolution, even though they generate the right propagators. 
This outcome relates to the basic properties of the integrals
over the even elements of a Grassmann algebra where the composites live, which
entails that the propagators are no longer the inverse of the wave operators.
\end{abstract}
\vfill
\hfill
\eject

\section{Introduction}
\label{sec:intro}

In the present paper we discuss an approach to deal with fermionic composites in
the framework of the field theory of elementary particles and of many-body
systems.  
This approach is
based on a change of  variables in the Berezin integrals defining the partition
function of the fermion system, whereby one introduces as integration variables
the composites one is interested in \cite{Palu}. Although the method is fairly
general, being suitable for describing trilinear as well as bilinear composites,
in this paper we confine ourselves to the latter case.

In the physics of elementary particles fermionic composites appear in general
as bound states, except when the vacuum is nontrivial due to fermion
condensation.  

In the domain of many-body systems bilinear composites are instead related 
to collective excitations which behave like bosons. A paradigmatic
example is of course provided by the Cooper pairs in metals \cite{Schr} and in
atomic nuclei \cite{Talm}, but other well known phenomena of bosonization are
met, e.g., in the density fluctuations of the electron gas \cite{Lieb}
(described by the Tomonaga model) and in the spin waves in
ferro-antiferromagnetic metals \cite{Marc}. 

As it is well known, the standard bosonization approach in the
theory of many-fermion systems is based on the so called Stratonovitch-Hubbard
\cite{Hub} representation of the partition function, where the fermion
action, quartic in the fermion  variables, is transformed, via the
introduction of an auxiliary bosonic field, into an action quadratic in the
fermion variables, which allows an exact integration over the latter to be
performed. If the resulting fermionic determinant can be expanded in a power
series, the kinetic terms needed to transform the auxiliary field into a
dynamical one can be generated. The latter field describes then
the collective excitations associated to the composite. A condition necessary 
for expanding the fermionic determinant is that these excitations should be
highly collective; however in general this is not sufficient, as we shall see in
a specific example. Moreover the method does not account for the Pauli
principle, which can be a serious shortcoming in atomic nuclei.
Apart from these limitations, the Stratonovitch-Hubbard transformation is
rather successful in the theory of many-body systems.

In the field theory of elementary particles, where the fermionic
terms are usually quadratic (quartic fermionic interactions being not
renormalizable in four dimensions), the Stratonovitch-Hubbard transformation is
of no use. In this context other methods have been proposed \cite{Kawa}, but
they appear to fail in providing a general solution to 
the problem \cite{Cara}: in particular none of them satisfactorily deals with
composites of low collectivity.

The approach discussed in the present paper proceeds through the following
steps:

i) the Grassmann variables associated to the composites are introduced as new
integration variables, 

ii) the free action for the composites is next determined,

iii) the composite propagators are then evaluated 

and, finally,

iv) the correlation functions in presence of an interaction are obtained.

This approach rearranges the usual perturbative
expansion in a hopefully advantageous way, because it describes the
propagation of the composites already at the zeroth
order. For illustrative purposes we will apply it to a simplified version of the
BCS theory, the so-called pairing model. The simplest version of the latter
considers a system of fermions living in an energy level of given angular
momentum $j$ and interacting pairwise only when the partners are coupled to zero
total angular momentum. Such a model is a
schematic representation of the physics of atomic nuclei far from closed shells,
where collective excitations associated with a spin zero composite occur. Its
spectrum is well known from the hamiltonian formalism
\cite{Rowe}: we derive it as well in the path integral formalism, which, as far
as we know, is a new result.

We now assess the status of the above program, outlining at the
same time the content of our paper. 

The first step has been accomplished in
refs.\cite{Palu}, whose results are summarized in Sec.\ref{sec:change} for the
ease of the reader. 

The second step has not yet been fully completed. Indeed first one has
to provide an unambiguous definition of the free action for the composites. 
Although the natural definition, when it exists, would appear to be the one
yielding the free propagators appropriate to the spin of the composite,
the calculation of the latters is far from trivial. For example for a scalar or
spinor composite, in order 
to achieve this goal one has to evaluate the correlation function of bilinear or
trilinear composites with an action which involves the forth or sixth power of
the fermionic variables. It is a notable feature of the present approach that
it allows the evaluation of the free propagators in a number of relevant cases,
which will be later described. Secondly a method has to be devised in order
to build the action of the composites out of the one of the constituents.

Clearly in relativistic field theories an action with the above outlined
features can be added by hands to the fermion one. Indeed 
in space-time dimensions greater than one it is 
a non-renormalizable, irrelevant operator, which modifies 
the renormalization group parameters by a finite amount.
But one should notice that, first,
this possibility does not occur in the framework of many-body systems, and, 
secondly, it would be clearly preferable to derive the action of the composites
from the original fermion one.
In Secs.\ref{sec:free} and \ref{sec:interacting} we shall present indeed one
way to achieve this scope for the simplified BCS model.

Let us now discuss the correlation functions. We already know [1] that
for trilinear, spin ${1 \over 2}$ composites, the free action is, under certain
conditions, the Dirac one, but the free action for scalars is still unknown.
A candidate, defined on a lattice, has been considered in ref. \cite{Palu}:
its propagator is related, in the continuum limit, to the 
self-avoiding random walk and lends itself, in a certain region of parameters,
to analytic approximations \cite{Cara}. Rather surprisingly the 
wave operator appearing in this action does not correspond to a true
time evolution. Yet a meaningful propagator can still be obtained because,
unlike in the case of fermions and true bosons, {\em for even composites the
propagator is not the inverse of the wave operator}, as we shall show in 
Sec.\ref{sec:change}.

In Secs.\ref{sec:free} and \ref{sec:interacting} of the present paper we derive
a free action for composites, whose wave operator again does not describe any
time evolution, although the associated propagators are the correct ones. In
this connection it is interesting to observe that the action of the composites
occurring in the pairing model has the same structure of the strong coupling
mesonic action of QCD \cite{Kawa,Klub}. This suggests that the wave
operator appearing in the latter case fails to describe the time evolution not
because of the nature of the strong
coupling, intrinsically far from the continuum, but rather because of a 
a general property of even composites.

In Sec.\ref{sec:interacting} we formulate as well the path integral for the
simplified BCS model in terms of the even variables associated to fermion pairs
with zero total angular momentum and we deduce the related spectrum. 
In Sec.\ref{sec:perturbation} we show how a two-body interaction can be handled
in a perturbative scheme. For simplicity we confine ourselves to the monopole
term of the multipole expansion of a general two-body force.

Finally in Sec.\ref{sec:concl} we present our conclusions emphasizing which
parts of the present application can be relevant to relativistic field
theories. 

\section{ Change of variables in a Berezin integral}
\label{sec:change}

Let us consider the composites

\begin{equation}
\Phi_I=\sum_{i_1,i_2=1}^{\Omega} k_I^{i_1i_2} \lambda_{i_1} \lambda_{i_2}
\label{eq:PhiI}
\end{equation}
where the $\lambda$ are the generating elements of a Grassmann algebra. The
variables $\Phi_I$ are characterized by the index of nilpotency $n^*(\Phi_I)$,
which is the smallest integer such that

\begin{equation}
\Phi_I^n=0\ \ \ \mbox{for}\ \ \ n>n^*(\Phi_I)\ .
\end{equation}

Our aim is to introduce the $\Phi$-variables as integration variables in
the Berezin integrals defining the composite correlation functions

\begin{equation}
<\Phi^*_{I_1}...\Phi_{I_n}>= {1 \over {Z_0}} \int [d\lambda^*d\lambda] 
\Phi^*_{I_1}...\Phi_{I_n}
e^{-S(\lambda^*,\lambda)}.
\end{equation}

In other words we must define an integral over the $\Phi$-variables in such a
way that, for an arbitrary function $f$,

\begin{equation}
\int[d\Phi]f(\Phi)=\int[d\lambda] f[\Phi(\lambda)] \ .
\end{equation}

Eqs.(\ref{eq:PhiI}) obviously cannot be inverted, namely the $\lambda$
cannot be expressed in terms of the $\Phi$. 
Nevertheless the latter can be considered as new
variables in the sense we specify below.

Indeed we first recall that if we have $2\Omega$ constituent variables only one
function (modulo a numerical factor) of the latter exists having a nonvanishing
Berezin integral. This function is the product of all the $\lambda$, namely

\begin{equation}
\Lambda= \lambda_1 \lambda_2 ... \lambda_{2\Omega}
\label{eq:Lambda}
\end{equation}
and its Berezin integral is

\begin{equation}
\int[d\lambda]\Lambda=1 \ , \ \ \mbox{with} \ \ \ [d\lambda]=d\lambda_{2\Omega} 
d\lambda_{2\Omega-1} 
... d\lambda_1\ , 
\end{equation}
all other integrals identically vanishing.
To find out the rules of integration over the variables $\Phi$, we must
therefore determine all the functions of the latter which, when expressed in
terms of the constituent variables, are proportional to $\Lambda$ with a nonzero
coefficient.  We call these functions {\em relevant}: in fact only when relevant
functions can be built out of the given composites can the $\Phi$ be introduced
as new variables of integration.  

Now, since the most general function of nilpotent variables is a polynomial,
it is sufficient to determine all the {\em relevant monomials}, namely the
monomials of maximum degree. These are referred to as $fundamental$ if they can
be expressed as products of powers of the $\Phi$ having the unity as a
coefficient and such that

\begin{equation}
\Theta_{\vec n}=\Phi_1^{n_1}\Phi_2^{n_2}...=w_{\vec n}\Lambda, \;\;\;w_{\vec
n}\neq 0 \ . 
\end{equation}

The components $n_I$ of the vector index ${\vec n}$, which characterizes these
monomials together with the weight $w_{\vec n}$, satisfy the constraint

\begin{equation}
\sum_I n_I =\Omega  \ .
\end{equation}

Notably, any integral over the variables $\Phi$ can be expressed in terms
of the weights. In fact a function of the $\Phi$ can always be expanded as
follows 

\begin{equation}
f(\Phi)=\sum_{\vec n} f_{\vec n} \Theta_{\vec n}+ \;\mbox{irrelevant terms}\ .
\end{equation}

If $f$ is expressed, via the definition of the $\Phi$, in terms of
the $\lambda$, then its Berezin integral is

\begin{equation}
\int [d\lambda]f[\Phi(\lambda)]=\sum_{\vec n} f_{\vec n}w_{\vec n}\ .
\end{equation}

This result holds as well when we integrate over the variables $\Phi$,
providing the rule of integration

\begin{equation}
\int[d\Phi]\Theta_{\vec n}=w_{\vec n}
\end{equation}
holds for the fundamental monomials, all the other integrals being zero.
Note that, although in general different expansions

\begin{equation}
f(\Phi)=\sum_{\vec n} f_{\vec n} \Theta_{\vec n}+\mbox{irr. terms}
=\sum_{\vec n} f'_{\vec n} \Theta_{\vec n} + \mbox{irr. terms}
\end{equation}
exist, the above equation implies

\begin{equation}
\sum_{\vec n} f_{\vec n} w_{\vec n}= \sum_{\vec n} f'_{\vec n}w_{\vec n},
\end{equation}
since both the l.h.s. and the r.h.s. are equal to the coefficient of $\Lambda$
in the expansion of $f(\Phi)$ in terms of the generating elements. Hence the
value of the integral does not depend upon the particular expression for $f$.

It should now be clear in which sense relations like (\ref{eq:PhiI}) can be
viewed as a change of variables, even though they cannot be inverted. In fact
in any nonvanishing Berezin integral the integrand must be proportional to
$\Lambda$, which can always be replaced by relevant functions of the composite
variables. 

This integration has some distinctive features:

1) the number of even variables needs not to be equal to half the number of the
   fermionic ones \cite{Cara}.

2) The integral of a fundamental monomial is unaffected by an arbitrary shift
   $\Phi \rightarrow \Phi+\alpha$.  
This property, however, does not apply to the integral of an arbitrary
function: in fact the expansion of the latter in fundamental monomials
changes as a consequence of the shift. Unlike the Berezin integral,
therefore, the integral over even elements of a Grassmann algebra is not
translationally invariant. Yet an equation of motion for
the even Grassmann fields can still be derived.

3) The integral over even Grassmann variables is not invariant under a
   transformation from even to even variables such to alter the index of
   nilpotency. This includes Fourier transformations. For this reason the
   propagator of even Grassmann variables cannot be evaluated in the standard
   way. 

4) The integral of the exponential of a quadratic form of variables of index 1
   is straightforward, namely

\begin{equation}
\int [d\Phi^* d\Phi] e^{\sum_{I,J} \Phi^*_I M_{I,J} \Phi_J}= per M\ , 
\end{equation}
where $perM$ is the permanent of the matrix $M$. 
This is the reason why the propagators for even Grassmann variables do not
coincide with the inverse of the wave operators $M_{I,J}$ and are difficult to
calculate. In fact the permanent of a matrix cannot be evaluated by
diagonalization, in turn reflecting the non-invariance 
of the integral under unitary transformations. 

In the present paper the fermionic variables $\lambda_m$ are  associated to
nucleons 
of given angular momentum $j$ and $z$-component $m$. We first introduce then the
even variables $\phi_m$ associated to nucleon pairs with zero component of
the angular momentum on the spin quantization axis 

\begin{equation}
\phi_m=\lambda_{-m}\lambda_m \ , \ \ \ \ m>0\ .
\end{equation}

Obviously they have index of nilpotency 1 and 
only one fundamental monomial of weight 1, viz. 

\begin{equation}
\Theta=\phi_{1/2}...\phi_j \ ,
\end{equation}
can be built out of them.

Next we consider the variable $\Phi$ expressed as a superposition of nucleon
pairs of zero total angular momentum according to
\footnote{The coupling of the two fermions to zero angular momentum would
actually entail a factor $(-1)^{j-m}/\sqrt{2j+1}$. In this case the pairing
interaction appearing in Eq.(\ref{eq:Sp}) would be multiplied by a factor
$\Omega$. We have assumed the definition (\ref{eq:A}) in order to simplify the
notations.}

\begin{equation}
\Phi=\sum_{m>0} (-1)^{j-m} \phi_m \ .
\label{eq:A}
\end{equation}
The index of nilpotency of $\Phi$ is $\Omega$ and only one fundamental monomial,
namely

\begin{equation}
\Theta=\Phi^{\Omega}\ ,
\end{equation}
is associated with $\Phi$: its weight is $\Omega!$.

To perform integrals over the even Grassmann variables when powers of the
variables $\Phi$ are present the following identities are helpful:

\begin{eqnarray}
\int [d\phi] \Phi^n=& &\delta(n-\Omega) \Omega!= \int d\Phi \Phi^n
\nonumber\\
\int[d\phi] \phi_m \Phi^n=& &\delta(n+1-\Omega)(\Omega-1)!= {1 \over
{\Omega}}\int d\Phi \Phi^{n+1}  
\nonumber\\
\int[d\phi]\phi_{n_1}...\phi_{n_k}\Phi^n=& &\delta(n+k-\Omega)
(\Omega-k)!= {1 \over {\Omega...(\Omega-k+1)}} \int d\Phi \Phi^{n+k}\ .
\end{eqnarray}

In fact from the above one sees that w.r. to the integration over $\Phi$, a
product of $\phi$ behaves as 

\begin{equation}
\phi_{n_1}...\phi_{n_k} \sim {(\Omega-k)! \over { \Omega !}} \Phi^k\ .
\label{eq:a1ak}
\end{equation}

\section{The free action and the free propagator of even fields of index 1}
\label{sec:free}

In this Section we introduce the free action for the $\phi$-variables and
evaluate their propagator. The partition function of a free system of fermions
reads 

\begin{equation}
Z_0= \int [d\lambda^*d\lambda]e^{-S_0}
\end{equation}
where

\begin{equation}
[d\lambda^*d\lambda]=\prod_m \prod _{t=1}^T d\lambda_m(t)^* d\lambda_m(t)
\end{equation}
is the volume element and

\begin{equation}
S_0=\tau \sum_{t=1}^T \sum_m \lambda^*_m(t)
(\nabla_t+e_m)\lambda_m(t-1)
\end{equation}
is the euclidean action.
In the above $\tau$ is the time spacing, $\nabla_t$ is the
discrete time derivative

\begin{equation}
\nabla_t f(t)=\frac{1}{\tau}[f(t+1)-f(t)]
\end{equation}
and $e_m$ is the fermion energy.
Note that the variables at the time $t=0$ are related to the ones at
$t=T$ by antiperiodic boundary conditions

\begin{equation}
\lambda_m(0)=-\lambda_m(T)\;.
\label{eq:antip}
\end{equation}

Let us now introduce the even elements of the Grassmann algebra, which describe
pairs of fermions with $J_z=0$

\begin{equation}
\phi_m(t)=\lambda_{-m}(t)\lambda_m(t),\;\;\;\phi^*_m(t)=\lambda^*_m(t)
\lambda^*_{-m}(t)
\label{eq:a}.
\end{equation}
We want to write the partition function as an integral over these variables

\begin{equation}
Z_0=\int [d\phi^*d\phi]e^{-S_0}\;.
\end{equation}

As a first step to express $S_0$ in terms of the $\phi$'s let us write the
free action as follows

\begin{eqnarray}
S_0 &=& \sum_{t=1}^T  \sum_{m>0} \left\{ \lambda^*_m(t)\lambda_m(t)+
\lambda^*_{-m}(t)\lambda_{-m}(t) 
\right.
\nonumber
\\
&-& \left. 
x_m\left[\lambda^*_m(t)\lambda_m(t-1)+\lambda^*_{-m}(t)\lambda_{-m}(t-1)\right]
\right\} 
\end{eqnarray}
where

\begin{equation}
x_m=1-\tau e_m\;.
\end{equation}

Next we express the exponential of (minus) the free action as the product
of exponentials (associated with different times and different absolute values
of $m$) and expand
each exponential accounting for the nilpotency of the variables.
We get

\begin{eqnarray}
e^{-S_0} &=&\prod_{t=1}^T \prod_{m>0} \left\{\left[1+\phi^*_m(t)
\phi_m(t)- 
(\lambda^*_m(t) \lambda_m(t)+\lambda^*_{-m}(t)\lambda_{-m}(t))\right]\right.
\\
& & \left.\left[1+x_m^2\phi^*_m(t)\phi_m(t-1)+x_m(\lambda^*_m(t)\lambda_m
(t-1)+ \lambda^*_{-m}(t)\lambda_{-m}(t-1))\right]\right\}
\nonumber
\end{eqnarray}
or, by explicitly performing the products,

\begin{eqnarray}
\label{eq:e-S0}
e^{-S_0} &=&
 \prod_{t=1}^T \prod_{m>0} \left\{1+\phi^*_m(t)\phi_m(t)+x_m^2 
\phi^*_m(t)\phi_m(t-1)\right.
\\
&+& x_m \phi^*_m(t)\left[\lambda_m(t)\lambda_{-m}(t-1) +
\lambda_m(t-1)\lambda_{-m}(t)\right]
\nonumber\\
&-& \left. \left[\lambda^*_m(t)\lambda_m(t)+\lambda^*_{-m}(t)\lambda_{-m}(t)
\right]
+x_m\left[\lambda^*_m(t)\lambda_m(t-1)+\lambda^*_{-m}(t)\lambda_{-m}(t-1)
\right]\right\}
\;.
\nonumber
\end{eqnarray}
Now one easily realizes that the terms in the last row of the
r.h.s. of (\ref{eq:e-S0}) {\it  contribute
neither to $Z_0$ nor to the expectation values of operators involving only the
$\phi_m$ and not the $\lambda_m$ variables}. Hence they can be neglected in
the evaluation of the propagator of the $\phi$.

In order to get the partition function, one has to identify in
(\ref{eq:e-S0}) the terms which generate monomials involving all the
$\phi^*$ and $\phi$'s. One obtains

\begin{eqnarray}
\int [d\phi^* d\phi] e^{-S_0}&=&\int [d\phi^* d\phi]
\prod_{m>0}\left\{ \prod_{t=1}^T \phi^*_m(t)\phi_m(t) + \prod_{t=1}^T
x^2_m\phi^*_m(t)\phi_m (t-1) \right. 
\nonumber\\
& &
\left. +\prod_{t=1}^Tx_m \phi^*_m(t) \left[\lambda_m(t)\lambda_{-m}(t-1)+
\lambda_m(t-1)\lambda_{-m}(t)\right]\right\} ,
\end{eqnarray}
where the integrals are straightforward and yield, in the limit
$T\rightarrow\infty$, the well known result

\begin{equation}
Z_0=\prod_{m>0}(1+x_m^T)^2 \tot0 \prod_{m>0} \left[1+e^{-\beta e_m}
\right]^2,\;\;\beta=\tau T. 
\end{equation}

Let us now evaluate the correlation function between two composite variables

\begin{equation}
<\phi_n(t_2) \phi^*_n(t_1)>=\frac{1}{Z_0} \int [d\phi^*d\phi]
\phi_n(t_2) \phi^*_n(t_1) e^{-S_0}. 
\end{equation}
Proceeding as before we find

\begin{equation}
<\phi_n(t_2)\phi^*_n(t_1)>=
\frac{1}{Z_0} x_n^{2(t_2-t_1)} \prod_{m\neq n} [1+x_m^T]^2
\tot0
\frac{e^{-(\beta_2-\beta_1)2e_n}}
{\left[1+e^{-\beta e_n }\right]^2}
\label{eq:corr12}
\end{equation}
for $t_1<t_2$ and

\begin{equation}
<\phi_n(t_2)\phi^*_n(t_1)>=
\frac{1}{Z_0} x_n^{2(t_2-t_1+T)} \prod_{m\neq n} [1+x_m^T]^2
\tot0
\frac{e^{-(\beta+\beta_2-\beta_1) 2e_n}}
{\left[1+e^{-\beta e_n}\right]^2}
\label{eq:corr21}
\end{equation}
for $t_1>t_2$.
Note that Eqs.(\ref{eq:corr12}) and (\ref{eq:corr21}) are just the squares
of the corresponding correlation functions for the $\lambda$ variables. However,
here the existence of a backward propagation is an artifact due to
antiperiodic boundary conditions. We must therefore take $\beta \gg
|\beta_2-\beta_1|$ in order to get rid of this artifact.

Note also that now the term linear in $x_m$ in 
Eq.(\ref{eq:e-S0}) does not contribute to the numerator.
Accordingly it will not affect the energy of the system and we can assume
as free action

\begin{equation}
S^{(e)}_0=-\sum_{t=1}^T \sum_{m>0} \left[\phi^*_m(t) \phi_m(t)
+x^2_m \phi^*_m(t) \phi_m(t-1) \right]\ .
\label{eq:e-S0a}
\end{equation}
 This action is quite different from the true boson and fermion
ones, {\em because in the continuum limit it does not contain any time
derivative}. As explained in the Introduction we conjecture that this is a
general feature of the free actions of even Grassmann fields.

Note that if the partition function is calculated with
(\ref{eq:e-S0a}), one gets a wrong result

\begin{equation}
Z_0=\prod_{m>0}(1+x_m^{2T}) \tot0 \prod_{m>0}\left[1+e^{-2\beta e_m
}\right] \;.
\end{equation}

We can therefore use $S_0^{(e)}$ to evaluate the spectrum and the ratios of
correlation functions, but not their absolute normalizations. 

Finally we should
emphasize that, although the variables $\phi$ are introduced to bosonize the
system, they are completely equivalent to the $\lambda$ and can 
be used therefore to evaluate also the propagator of a {\em single nucleon}, as
shown in the Appendix.

\section{The pairing interaction}
\label{sec:interacting}

In this Section we consider a system of fermions interacting via a pairing
potential, whose action reads

\begin{equation}
S = \sum_{t=1}^T
\left\{\sum_{m>0} \left[-\phi^*_m(t) \phi_m(t) - x^2 \phi^*_m(t)
\phi_m (t-1)\right]  
- g_{_P}\tau \Phi^*(t) \Phi(t-1)\right\}
\label{eq:Sp},
\end{equation}
where

\begin{equation}
\Phi(t)=\sum_{m>0} (-1)^{j-m} \lambda_{-m}(t) \lambda_m(t)
\end{equation}
describes a pair of fermions with zero total angular momentum. We have
included only the term $S_0^{(e)}$ from the free fermionic action, because 
the other terms cannot contribute to correlation functions of the $\Phi$.
Moreover we deal for simplicity only with the degenerate case, i.e. 
$e_m=e, x_m=x$.

The spectrum of the system described by the action (\ref{eq:Sp}), restricted to
states of $n$ spin zero pairs of nucleons (the so-called seniority zero states),
is \cite{Rowe} 
\begin{equation}
E_n=n[2e-(\Omega-n+1)g_{_P}]
\label{eq:spectrum}
\end{equation}
and it is characterized by the parameter

\begin{equation}
\xi={(\Omega+1)g_{_P} \over {2e}}\ .
\end{equation}
We will confine ourselves to the case $\xi<1$, since then the ground state is
trivial. Worth mentioning is that, even in presence of high collectivity,
insured by a large value of $\Omega$, the fermionic determinant arising after 
the application of the Stratonovitch-Hubbard
transformation can be expanded in a power series \cite{Cara} only if $\xi\ll1$. 
Our method is not affected by such a restriction.

The correlation functions of $n$ spin zero pairs of nucleons are 

\begin{equation}
< \Phi^n(t_2) \left[\Phi^*(t_1)\right]^n >
= {1 \over Z} \int [d\phi^*d\phi] \Phi^n(t_2) \left[\Phi^*(t_1)\right]^n
e^{-S}  \ .
\end{equation}

Their evaluation is simplified by introducing the variables

\begin{eqnarray}
{\cal B}(t) & \equiv & \sum_{m>0} \phi_m(t)^* \left[\phi_m(t) 
+ x^2 \phi_m(t-1) \right]
\nonumber\\
& \equiv & \sum_{m>0} {\cal C}_m(t) ,
\end{eqnarray}
in terms of which
\begin{eqnarray}
S &=& \sum_{t=1}^T \left[-{\cal B}(t) - g_{_P}\tau \Phi^*(t) \Phi(t-1)\right]
\nonumber
\\
&\equiv& \sum_{t=1}^T S(t)
\end{eqnarray}
and
\begin{eqnarray}
\label{eq:ncorr}
& &< \Phi^n(t_2) \left[\Phi^*(t_1)\right]^n >
= { 1 \over Z}\int [d\phi^*d\phi] \Phi^n(t_2) \left[\Phi^*(t_1)\right]^n 
\prod_{t=1}^T\sum_{r=0}^{\Omega} \frac{\left[-S(t)\right]^r}{r!} \\
& = & \int [d\phi^*d\phi] \Phi^n(t_2) \left[\Phi^*(t_1)\right]^n
\prod_{t=1}^T 
\sum_{r=0}^{\Omega} \frac{1}{r!} \sum_{k=0}^r {r\choose k} {\cal B}^{r-k}(t)
(g_{_P}\tau)^k \left[\Phi^*(t)\Phi(t-1)\right]^k \;.
\nonumber
\end{eqnarray}

The powers of ${\cal B}$ appearing in (\ref{eq:ncorr}) are given by

\begin{eqnarray}
{\cal B}^r(t) & = & \left[\sum_{m>0} {\cal C}_m(t)\right]^r
\nonumber \\
& = & \sum_{k_1+...k_\Omega=r} \frac{r!}{k_1!...k_\Omega!} {\cal
C}_1^{k_1}(t)... {\cal C}_\Omega^{k_\Omega}(t)
\nonumber\\
& = & r!\sum_{k_1+...k_\Omega=r} {\cal C}_1^{k_1}(t)...{\cal
C}_\Omega^{k_\Omega}(t),\;\;\;\;\;\;k_i=0,1
\label{eq:Br1}
\end{eqnarray}
where the restriction on the indices $k$ stems from the nilpotency of the
variables ${\cal C}$.

Now each of the ${\Omega\choose r}$ terms (the number of combinations of class
$r$ of $\Omega$ objects) in the r.h.s. of Eq.(\ref{eq:Br1}) provides
the same contribution. Accordingly in the integral ${\cal B}^r$ behaves as

\begin{eqnarray}
{\cal B}^r(t)
& & \sim r! {\Omega \choose r} {\cal C}_1(t)...{\cal C}_r(t) \\
& & \sim r! {\Omega \choose r} \phi^*_1(t)...\phi^*_r(t)
\sum_{s=0}^r{r\choose s} \phi_1(t)...\phi_s(t) x^{2(r-s)} 
\phi_{s+1}(t-1)...\phi_r(t-1) 
\nonumber
\end{eqnarray}
due to the symmetry w.r. to permutations of the $\phi$'s. Because of this
symmetry, as far as the integration over the $\Phi$ is concerned, the rule 
(\ref{eq:a1ak}) holds, so that
\begin{equation}
{\cal B}^r(t)
\sim \frac{1}{(\Omega!)^2} \left[\Phi^*(t)\right]^r\sum_{s=0}^r {r\choose s}
(\Omega-s)! \left[\Omega-(r-s)\right]! x^{2(r-s)} \Phi^s(t) \Phi^{r-s}(t-1) \;.
\label{eq:Br}
\end{equation}
Now, in relation to the integration over $\Phi^*$, the correlation
function is non-vanishing if and only if a power $\left[\Phi^*(t)\right]^\Omega$
appears in the integrand for each time $t$. Therefore (\ref{eq:ncorr}) will not
vanish only when
\begin{eqnarray}
 r=\Omega   & \mbox{for}& t \ne t_1 \nonumber\\
 r=\Omega-n & \mbox{for}& t  =  t_1 .
\end{eqnarray}
On the other hand, in the integration over the variable $\Phi$ at the time
$t=t_2$, in order to recover the power $\Phi^\Omega(t_2)$, one should single out
from ${\cal B}^{\Omega-k}(t_2)$ a factor $\Phi^{\Omega-n}(t_2)$. It is then an
easy matter to obtain

\begin{equation}
\int d\Phi^*(t_2) d\Phi(t_2) \Phi^n(t_2) \frac{\left[-S(t_2)\right]
^\Omega}{\Omega!} =\alpha \Phi^n(t_2-1)\ ,
\end{equation}
where $\alpha$ is a number which will be evaluated below. This
shows that the integral over $\Phi^n(t_2)$ is proportional to $\Phi^n(t_2-1)$,
so that, by performing the integrals in the range $t_1<t\leq t_2$ we get

\begin{equation}
\int \left[d\Phi^*(t) d\Phi(t)\right]_{t_1<t\leq t_2} \Phi^n(t_2)
\prod_{t=t_1+1}^{t_2} \frac{\left[-S(t)\right]^\Omega}{\Omega!}
= \alpha^{t_2-t_1} \Phi^n(t_1) \;.
\end{equation}
In the limit of vanishing $\tau$ one has
\begin{equation}
\lim_{\tau\to 0} \alpha^{t_2 - t_1} =
\lim_{\tau\to 0} \alpha^{(\hat t_2 -\hat t_1)/\tau} =
\lim_{\tau\to 0} e^{(\hat t_2 -\hat t_1)/\tau \ln\alpha} =
e^{-E(\hat t_2 -\hat t_1)} \;,
\end{equation}
where the physical time $\hat t = \tau t$ has been introduced and 

\begin{equation}
E=\lim_{\tau\to 0} \left(-\frac{1}{\tau} \ln\alpha\right)
\end{equation}
has the significance of the energy. The coefficient $\alpha$ needs to be
calculated to first order in $\tau$ only. In turn this entails
\begin{eqnarray}
\frac{\left[-S(t)\right]^\Omega }{\Omega!}
& = & \frac{1}{\Omega!} \sum_{k=0}^\Omega {\Omega\choose k} {\cal
B}^{\Omega-k}(t)(g_{_P}\tau)^k \left[\Phi^*(t) \Phi(t-1)\right]^k \nonumber\\
& \simt0 & \frac{1}{\Omega!} \left[ {\cal B}^\Omega(t) + \Omega
g_{_P}\tau {\cal B}^{\Omega-1}(t) \Phi^*(t) \Phi(t-1)\right] .
\label{eq:SOm}
\end{eqnarray}
By inserting (\ref{eq:Br}) into (\ref{eq:SOm}) one then gets
\begin{eqnarray}
\frac{\left[-S(t)\right]^\Omega}{\Omega!} &\simt0&
\frac{\left[\Phi^*(t)\right]^\Omega}{(\Omega!)^2}  \{ \Phi^\Omega(t) 
\label{eq:Som1}
\\
& + &
\sum_{s=0}^{\Omega-1} \left[ x^{2(\Omega-s)} + (s+1) (\Omega-s)
g_{_P}\tau\right] \Phi^s(t) \Phi^{\Omega-s}(t-1)\} 
\nonumber
\end{eqnarray}
and since
\begin{equation}
x^{2(\Omega-s)} \simt0 1-2(\Omega-s) e\tau
\label{eq:x2}
\end{equation}
one can finally cast (\ref{eq:Som1}) in the form
\begin{eqnarray}
\frac{\left[-S(t)\right]^\Omega}{\Omega!}
& = & \frac{\left[\Phi^*(t)\right]^\Omega}{(\Omega!)^2} \{ \Phi^\Omega(t)
\\
& + &
\sum_{s=0}^{\Omega-1} \left[ 1-\left(2(\Omega-s)e
- (s+1)(\Omega-s)g_{_P}\right)\tau\right] \Phi^s(t) \Phi^{\Omega-s}(t-1)\}
\;. 
\nonumber
\end{eqnarray}
Therefore for the relevant integral appearing in the correlation function one
obtains
\begin{equation}
\int d\Phi^*(t_2) d\Phi(t_2) \Phi^n(t_2) \frac{\left[-S(t_2)\right]
^\Omega}{\Omega!}\simt0 \Phi^n(t_2-1)\left\{1-n\left[2e-(\Omega-n+1)
g_{_P}\right] \tau\right\},
\end{equation}
from where it follows

\begin{equation}
\alpha=1-[2e - ( \Omega-n+1)g_{_P}]\tau n
\end{equation}
and 
\begin{equation}
< \Phi^n(t_2) \left[\Phi^*(t_1)\right]^n > \simt0 
e^{-(\beta_2-\beta_1)[2e-(\Omega-n+1)g_{_P}]n}
\end{equation}
for the correlation function. 
Hence for the energy spectrum of the system one recovers the well known
formula (41).

From the above calculation the free action for the field
$\Phi$ is easily deduced. In fact, setting $g_{_P}$=0 in Eq.(\ref{eq:ncorr}),
we see that this action can be written as

\begin{equation}
S_0^{(\Phi)} = - \sum_{t=1}^T \ln \left\{ 
\sum_{r=0}^{\Omega} \frac{1}{r!} \left[\Phi^*(t)\right]^r
\sum_{s=0}^r {r\choose s} 
\frac{(\Omega-s)! \left[\Omega-(r-s)\right]!}{(\Omega!)^2} 
x^{2(r-s)} \Phi^s(t) \Phi^{r-s}(t-1) 
\right\}\ .
\end{equation}

Clearly the free correlation functions, which turn out to be independent of the
total angular momentum of the pair, are then
\begin{equation}
< \Phi^n(t_2) \left[\Phi^*(t_1)\right]^n > \simt0 
e^{-(\beta_2-\beta_1)2en}\ .
\end{equation}

We have evaluated the logarithm for a few values of $\Omega$, always obtaining 
an action of the form
\begin{equation}
S_0^{(\Phi)} = - \sum_{t=1}^T \sum_{s=1}^\Omega \alpha_s 
\left[\Phi^*(t)\right]^s \left[ \Phi^s(t) + x^{2s} \Phi^s(t-1) \right]\ ,
\end{equation}
$\alpha_s$ being a numerical coefficient. It is indeed remarkable
that $S_0^{(\Phi)}$ displays the same structure of the mesonic action appearing
in the strong coupling limit of QCD \cite{Kawa,Klub}. Such a structure is
characterized by the absence of terms of the type 
$ [\Phi^*(t)]^n [\Phi(t)]^{n-s} [\Phi(t-1)]^s $ ($s\neq 0$), due to the
occurrence of nontrivial cancellations.

Obviously the total action can be recast in the form
\begin{equation}
S^{(\Phi)} = S_0^{(\Phi)} - g_{_P}\tau\sum_{t=1}^T\Phi^*(t)\Phi(t-1)\ .
\end{equation}

\section{Perturbation theory}
\label{sec:perturbation}

In this Section we show how interactions can be handled in a perturbative
scheme considering, as an example, the monopole term of the multipole expansion
of a quartic interaction. Here the quartic character of the interaction is
irrelevant, a similar calculation could be performed for a quadratic one, and
higher multipolarities can be dealt with in a similar fashion. 
The associated action can be written as follows

\begin{equation}
S_I =
-\frac{\tau R}{2} \sum_{t=1}^T \sum_{m\neq m'} \lambda^*_m(t) \lambda^*_{m'}(t) 
\lambda_{m'}(t-1) \lambda_m(t-1)\ ,
\label{eq:mono}
\end{equation}
where $R$ is the radial integral of the monopole term of the interaction,
and can be trivially split into two pieces: one expressed only in terms of
the $\phi$ variables and the other in terms of the odd Grassmann variables
$\lambda$ according to

\begin{equation}
S_I = S_1+S_2\ ,
\end{equation}
where
\begin{equation}
S_1 =
\tau R \sum_{t=1}^T \sum_{m>0} \phi^*_m(t) \phi_m(t-1) \ 
\end{equation}
and
\begin{equation}
S_2 =
-\frac{\tau R}{2} \sum_{t=1}^T \sum_{m} \sum_{m'\neq m} 
\lambda^*_m(t) \lambda^*_{m'}(t) \lambda_{m'}(t-1) \lambda_m(t-1)\ .
\end{equation}

For the partition function of the fermion system we shall accordingly have

\begin{equation}
Z=\int [d\lambda^* d\lambda] e^{-S}
\end{equation}
with
\begin{equation}
S=S_0+S_I.
\end{equation}

Now in general the terms in the third row of 
Eq.(\ref{eq:e-S0}), which expresses $S_0$, can no longer be neglected 
{\em because the $\phi$-variables could be reconstructed utilizing terms arising
from the expansion wr to $S_I$.}  

Thus, in order to be able to exponentiate the whole expression appearing in 
Eq.(\ref{eq:e-S0}), we introduce the even variable $\epsilon_m(t)$ of index 1
and rewrite the exponential of (minus) the action of the free fermions in the
form 

\begin{equation}
e^{-S_0} \rightarrow e^{-S_0^{(e)}-\bigtriangleup S_0}\ ,
\end{equation}
where

\begin{eqnarray}
\bigtriangleup S_0 &=& \sum_{t,m>0}
\epsilon_m(t)\left\{ \lambda^*_m(t)\left[ \lambda_m(t)-x \lambda_m(t-1)\right]
 + \lambda^*_{-m}(t)\left[ \lambda_{-m}(t)-x \lambda_{-m}(t-1)\right]\right\}
\nonumber\\
 & & -x \phi^*_m(t)\left[ \lambda_m(t) \lambda_{-m}(t-1)+\lambda_m(t-1) 
\lambda_{-m}(t)\right]\ ,
\end{eqnarray}
setting $\epsilon=1$ at the end of the calculations. 
$\bigtriangleup S_0$ should now be treated as a perturbation together 
with $S_I$. 
In each term of the perturbative expansion the $\lambda$'s must be combined into
the $\phi$'s, the integration over the $\phi$ must be performed and,
finally, the $\epsilon$ must be set equal to 1. 
It remains to be seen whether the expansion wr to $\bigtriangleup S_0$ is
permissible in the absence of a small parameter.

In the example we are considering we have, in first order of
perturbation theory, 
\begin{equation}
e^{-S} \simeq e^{-S_0^{(e)}} e^{-\bigtriangleup S_0} (1-S_1-S_2) \ .
\end{equation}
The related partition function is thus
\begin{equation}
Z = Z_0^{(e)} + Z_1 + Z_2\ ,
\label{eq:ZI}
\end{equation}
where
\begin{equation}
Z_0^{(e)} = (1+x^{2T})^{\Omega} \tot0 \left[1+e^{-2\beta e}\right]^{\Omega}\ 
\end{equation}
and the correlation function of the $\phi_m$,
which turns out to be independent of $m$, reads

\begin{equation}
< \phi_m^*(t_1) \phi_m(t_2) > = \frac{1}{Z} \left[ {\cal G}_0^{(1)}(t_1,
t_2) +  {\cal G}_1^{(1)}(t_1,t_2) + {\cal G}_2^{(1)}(t_1,t_2)\right]\ ,
\label{eq:GI}
\end{equation}
${\cal G}_0^{(1)}(t_1,t_2)/Z_0$ being the free pair propagator given by
Eqs.(\ref{eq:corr12}) and (\ref{eq:corr21}). 

Now we should first notice that odd powers of $\bigtriangleup S_0$ never come
into play: indeed the related Berezin integral vanishes owing to the
impossibility of reconstructing a complete product of the $\phi$ variables. 
It can moreover be checked that, for the specific interaction we are
considering, (\ref{eq:ZI}) and (\ref{eq:GI}) are not contributed to by 
powers of $\bigtriangleup S_0$ larger than two owing to the nilpotency
of the variables $\phi$ and $\epsilon$. Thus, in addition to $S_I$, only the
term quadratic in $\bigtriangleup S_0$ has to be considered, and this 
justifies the expansion of the exponential wr to $\bigtriangleup S_0$.

Concerning $S_1$ one gets
\begin{equation}
Z_1 = -\tau T R\Omega (1+x^{2T})^{\Omega-1} {x^{2(T-1)}}
\end{equation}
for the partition function and
\begin{equation}
{\cal G}_1^{(1)}(t_1,t_2) = 
-\tau R x^{2(t_2-t_1)} x^{-2} (1+x^{2T})^{\Omega-1} \left[ 
(t_2-t_1) + T (\Omega-1) \frac{x^{2T}}{1+x^{2T}} \right]
\end{equation}
for the correlation function when $t_1<t_2$.

More delicate is the analysis of the last terms in (\ref{eq:ZI}) and 
(\ref{eq:GI}), which brings into play 
$\bigtriangleup S_0$ via the coupling to $S_2$. For these it is indeed possible
to reconstruct the complete chain of $\phi$'s; one thus obtains

\begin{equation}
Z_2 = -2\tau T R\Omega (\Omega-1) (1+x^{2T})^{\Omega-2} {x^{2(2T-1)}}
\end{equation}
and
\begin{equation}
{\cal G}_2^{(1)}(t_1,t_2) = 
-4\tau R (\Omega-1) x^{2(t_2-t_1)} x^{2(T-1)} (1+x^{2T})^{\Omega-2} \left[ 
(t_2-t_1) + \frac{T}{2} (\Omega-2) \frac{x^{2T}}{1+x^{2T}} \right]\ .
\end{equation}

However, in the above expressions the terms proportional to 
$x^{2T}\sim e^{-2\beta e}$ provide a negligible contribution, according to the
discussion following Eq.(\ref{eq:corr21}).
As a consequence one simply gets for the pair correlation function in 
presence of the interaction (\ref{eq:mono}) and for $t_1<t_2$ the expression

\begin{equation}
<\phi^*_m(t_1) \phi_m(t_2)>
\tot0
\frac{e^{-(\beta_2-\beta_1)(2e+R)}}
{1+e^{-2\beta e}}\ ,
\label{eq:shift}
\end{equation}
which corresponds to a constant shift of the pair energy $2e\to 2e+R$. 

If we now consider the correlation function of two pairs, then the term $S_2$
{\em does} contribute through its coupling to $\bigtriangleup S_0$, yielding:

\begin{equation}
< \phi^*_m(t_1) \phi^*_n(t_1) \phi_m(t_2) \phi_n(t_2) > =
\frac{1}{Z}  \left[ {\cal G}_0^{(2)}(t_1,t_2) + 
{\cal G}_1^{(2)}(t_1,t_2) + 
{\cal G}_2^{(2)}(t_1,t_2)\right]\ .
\end{equation}
In the above, if $t_1<t_2$ and again neglecting the contributions 
proportional to $x^{2T}$,
\begin{equation}
{\cal G}_0^{(2)}(t_1,t_2) = (1+x^{2T})^{\Omega-2} x^{4(t_2-t_1)}
\end{equation}
\begin{equation}
{\cal G}_1^{(2)}(t_1,t_2) = -2R\tau (t_2-t_1) (1+x^{2T})^{\Omega-2} 
x^{4(t_2-t_1)} x^{-2}
\end{equation}
and
\begin{equation}
{\cal G}_2^{(2)}(t_1,t_2) = 2 {\cal G}_1^{(2)}(t_1,t_2) \ .
\end{equation}
By taking the limit $\tau\to 0$ one finally obtains 
\begin{equation}
< \phi^*_m(t_1) \phi^*_n(t_1) \phi_m(t_2) \phi_n(t_2) > 
\tot0
\frac{e^{-(\beta_2-\beta_1)2(2e+3R)}}
{(1+e^{-2\beta e})^2}\ ,
\end{equation}
which amounts to the shift $e\to e+\frac{3}{2}R$ for the single particle energy
with respect to the non-interacting case.

From the above results the perturbative corrections to the
spectrum of a system of one and two spin-zero pairs follow.
Indeed we have for the correlation function of one spin-zero pair
\begin{equation} 
< \Phi^*(t_1) \Phi(t_2) >
= \sum_{m_1,m_2>0} (-1)^{2j-m_1-m_2} <\phi^*_{m_1}(t_1) \phi_{m_2}(t_2) >\ ,
\end{equation} 
which, the correlator of the $\phi_m$ being diagonal in $m$ and
$m$-independent, entails 
\begin{equation} 
< \Phi^*(t_1) \Phi(t_2) >
= \Omega <\phi^*_m(t_1) \phi_m(t_2) > \ .
\end{equation} 

By a simple extension the correlator 
\begin{equation} 
< \left[\Phi^*(t_1)\right]^2 \left[\Phi(t_2)\right]^2 >
= 2\Omega(\Omega-1) 
<\phi^*_m(t_1) \phi^*_n(t_1) \phi_m(t_2) \phi_n(t_2) > 
\end{equation} 
is also obtained.

Clearly this scheme can be easily generalized to obtain the correlation
functions when both the monopole and the pairing interaction are acting.

\section{Conclusions}
\label{sec:concl}

In this paper we have tested a new approach to bosonization to be used in the
framework of the theory of relativistic fields and of many-body systems. A
new approach seems to be required especially in the context of quantum field
theory, where the Stratonovitch-Hubbard method does not help.
The present scheme, hopefully, might indeed offer some advantages because first
it describes correctly, already at the zeroth order, the propagation of
composites and furthermore because the introduction of the composites as
integration variables is expected, on general ground, to be helpful in fastening
the convergence of the perturbative expansion through a rearrangement of the
latter. 

The formalism utilized in the present work
has been developed in previous papers where the free
propagators have also been evaluated in a number of other relevant cases. 
The problem still waiting for a general solution is how to derive the free
action of the composites starting from the one of the constituents. In order to
explore this point and to study additional relevant 
features of the method, we have studied a simplified version of
the BCS model. 
In this simple context it has been possible to fully work out the new
bosonization scheme, obtaining the spectrum, already known in the hamiltonian
framework, in the path integral formalism.

In particular we have been able to derive the exact free action
for the composite starting from the one of the constituents. This is the first
example, to our knowledge, where such a derivation has been performed without
recurring to approximations.
It is presently unclear whether our procedure can be generalized to 
relativistic field theories or to many-body systems of greater complexity than
the degenerate pairing model. 
We however believe that the structure of our free action is not restricted to
the spin-zero composite of the pairing model, but is likely to be shared by
the actions of all even composites, in particular as far as the wave operator
is concerned. A support to this conjecture is provided by the form of the action
proposed in  ref. \cite{Palu} and also by the mesonic action in the strong
coupling limit of QCD. 

Finally a comment concerning the role played by the pieces of the free action
still expressed through the original odd Grassmann variables is in order. Terms
of this kind are indeed expected to be present in any realistic problem and, as
a rule, they occur without a small parameter in front. In this connection it is
of significance our finding in the context of the pairing model: here these
terms can only be effective when an interaction is acting among the constituents
and, accordingly, they can be treated as a perturbation together with the
latter. 
Also in this case we are inclined to believe this to be a general feature
of all the actions expressed through the even elements of a Grassmann algebra.

In conclusion the present study shows that the approach to deal with
composite fields, originally developed in ref.\cite{Palu}, can be successfully
applied to nonrelativistic many-body problems whenever the introduction of
complex building blocks (the composites) is hinted by the nature of the system.
In particular, and importantly, the method allows the derivation of the free
action for the composites and appears worth to be explored in the context of
more realistic situations.

\appendix
\section*{Appendix}

In this Appendix we want to show that, since the introduction of the $\phi$
is a change of variables, we can use them also to evaluate the propagator of a
single nucleon 

\begin{equation}
<\lambda_n(t_2) \lambda^*_n(t_1)>=\frac{1}{Z_0} \int [d\phi^*d\phi]
\lambda_n(t_2) \lambda^*_n(t_1)e^{-S_0}\ .
\label{eq:ccstar}
\end{equation}

To this purpose let us start by
considering the case $t_1<t_2$ and $n>0$. 
We shall take into account only the term $|m|=|n|$ in
(\ref{eq:e-S0}) since the other terms are canceled out by corresponding terms
of $Z_0$ in the denominator.
In order to construct $\phi^*_n(t_1)$, we must combine $\lambda^*_n(t_1)$
with the terms containing $\lambda^*_{-n}(t)$ in the $t=t_1$ term of the third
row of  (\ref{eq:e-S0}), which yields

\begin{equation}
\phi^*_n(t_1) [-\lambda_{-n}(t_1) + x_n \lambda_{-n}(t_1-1)] \ .
\end{equation}
The two terms of the above expression are the only insertions arising
from the third row of (\ref{eq:e-S0}) which, when combined with the terms in
the first row at $t\ne t_1$, give rise to two ``chains'' containing all the 
$\phi$ and $\phi^*$ and thus yielding a non-zero Berezin integral. The
first chain is (a minus sign comes from the interchange of $\lambda^*_n(t_1)$
and $\lambda_n(t_2)$)

\begin{eqnarray}
&-&[\phi^*_n(1)\phi_n(1)] \ ...\  [\phi^*_n(t_1-1)\phi_n(t_1-1)]
[-\phi^*_n(t_1)\lambda_{-n}(t_1)] 
\nonumber\\
&\times&
[x_n\phi^*_n(t_1+1)\lambda_n(t_1)\lambda_{-n}(t_1+1)] \ ... \ 
[x_n\phi^*_n(t_2)\lambda_n(t_2-1)\lambda_{-n}(t_2)]
\lambda_n(t_2) \nonumber\\
&\times&
[\phi^*_n(t_2+1)\phi_n(t_2+1)] \ ...\  [\phi^*_n(T)\phi_n(T)]
\nonumber\\ 
&=& x_n^{t_2-t_1} \prod_{t=1}^T \phi^*_n(t) \phi_n(t) \ ,
\end{eqnarray}
while the second one reads

\begin{eqnarray}
&-&(x_n\phi^*_{n,1}\lambda_{n,1}\lambda_{-n,0})\ ...\ [x_n\phi^*_n(t_1-1)
\lambda_n(t_1-1)\lambda_{-n}(t_1-2)]
[x_n\phi^*_n(t_1)\lambda_{-n}(t_1-1)] \nonumber\\
&\times&
[x_n^2\phi^*_n(t_1+1)\phi_n(t_1)] \ ...\  (x_n^2\phi^*_n(t_2)
\phi_n(t_2-1)] \lambda_n(t_2) \nonumber\\
&\times&
[x_n\phi^*_n(t_2+1)\lambda_n(t_2+1)\lambda_{-n}(t_2)] \ ...\  
[x_n\phi^*_n(T)\lambda_n(T)\lambda_{-n}(T-1)]
\nonumber\\
&=& x_n^{t_1} x_n^{2(t_2-t_1)} x_n^{T-t_2} \prod_{t=1}^T \phi^*_n(t)
\phi_n(t) 
= x_n^{t_2-t_1+T} \prod_{t=1}^T \phi^*_n(t) \phi_n(t) \ ,
\end{eqnarray}
where the antiperiodicity condition (\ref{eq:antip}) has been used.
Therefore

\begin{equation}
<\lambda_n(t_2) \lambda^*_n(t_1)>=
\frac{1}{Z_0} x_n^{t_2-t_1} (1+x_n^T) \prod_{|m|\neq |n|} [1+x_m^T]^2
\tot0
\frac {e^{-(\beta_2-\beta_1)e_n}} {1+e^{-\beta e_n}}\ .
\end{equation}
Similarly, for $t_1>t_2$ one gets 
\begin{equation}
<\lambda_n(t_2) \lambda^*_n(t_1)>=
\frac{1}{Z_0} x_n^{T-t_1+t_2} (1+x_n^T) \prod_{|m|\neq |n|} [1+x_m^T]^2
\tot0
- \frac {e^{-(\beta_2-\beta_1)e_n}} {1+e^{\beta e_n}}\ .
\end{equation}

\end{document}